\documentclass[useASM]{mn2e}
\input{psfig.sty}
\usepackage{graphicx}
\begin{document}
\title{Testing a new luminosity/redshift indicator for $\gamma$-ray bursts}
\date{2007 Apr. 13}
\pubyear{????} \volume{????} \pagerange{4} \twocolumn

\author[Zhang \& Xie]
       {Zhi-Bin Zhang$^{1, 2}$ and Guang-Zhong Xie$^{2}$\\
$^1$International Center for Astrophysics, Korea Astronomy and Space
Science Institute, 36-1 Hwaam, Yusong, Daejon
305-348,\\ South Korea; Email: zbzhang@kasi.re.kr\\
$^2$Yunnan Observatory, National Astronomical Observatories, Chinese
Academy of Sciences, P. O. Box 110, Kunming 650011,\\ China}
\date{}

\pagerange{\pageref{firstpage}--\pageref{lastpage}} \pubyear{2007}

\maketitle
\label{firstpage}
\begin{abstract}

We have tested a relative spectral lag (RSL) method suggested
earlier as a luminosity/redshift (or distance) estimator, using the
generalized method by Schaefer \& Collazzi. We find the derivations
from the luminosity/redshift-RSL (L/R-RSL) relation are comparable
with the corresponding observations. Applying the luminosity-RSL
relation to two different GRB samples, we find that there exist no
violators from the generalized test, namely the Nakar \& Piran test
and Li test. We also find that about 36 per cent of Schaefer's
sample are outliers for the L/R-RSL relation within 1$\sigma$
confidence level, but no violators at 3$\sigma$ level within the
current precision of L/R-RSL relation. An analysis of several
potential outliers for other luminosity relations shows they can
match the L/R-RSL relation well within an acceptable uncertainty.
All the coincident results seem to suggest that this relation could
be a potential tool for cosmological study.

\end{abstract}

\section{Introduction}

Gamma-ray bursts (GRBs) have been found to have several empirical
correlations based on the properties of light curves and spectra
(e.g., Norris, Marani \& Bonnell 2000; Lloyd, Petrosian \& Mallozzi
2000; Fenimore \& Ramirez-Ruiz 2000; Chang et al. 2002; Schaefer
2003a; Amati et al. 2002; Ghirlanda et al. 2004; Schaefer 2002;
Firmani et al. 2006). It has been proposed that the relations can be
used as ``standard candle'' for cosmological applications (e.g.
Ghirlanda et al. 2006 and Schaefer 2007 for reviews; see also
Friedman \& Bloom 2005; Bertolami \& Silva 2006; Oguri \& Takahashi
2006; Li 2007b) thanks to the feature of the standard energy release
from the central engine (Frail et al. 2001; Bloom et al. 2003).
However, as pointed by Schaefer \& Collazzi (2007), these empirical
relations can return the luminosities with a diverse level of
accuracy. That is, the derived luminosity is highly dependent on the
estimator or relation. It is therefore need to inspect these
estimators by comparing the derived and observed luminosities
directly.

Following the tests of Nakar \& Piran (2005) and Li (2007a),
Schaefer \& Collazzi (2007) gave a generalized test (Section 2)
simultaneously to the eight luminosity relations. As a result, all
the eight luminosity relations passed the generalized test
successfully. For the $E_{p}-E_{\gamma, iso}$ relation (Amati et al.
2002), they found that $\sim44\%$ of 69 GRBs in their sample
including pre-Swift and Swift sources were violators. However, they
explained that this was a natural consequence resulting from both
systematic and observational errors (due to the small fluctuations
of $<1\sigma$ deviations). By analyzing a BATSE sample, Band \&
Preece (2005) otherwise found $\sim88\%$ of pre-Swift bursts were
inconsistent with the $E_{p}-E_{\gamma, iso}$ relation. Further
investigations (Cabrera et al. 2007; Butler et al. 2007) show the
``Amati relation'' does exist in the Swift sample but is already
inconsistent with its pre-Swift form (Lamb 2004). Recently, Butler
et al. (2007) systematically investigated the properties of temporal
and spectral parameters for 218 Swift bursts and pointed out the
$E_{p}-E_{\gamma, iso}$ relation, as well as most other pre-Swift
relations proposed by Yonetoku et al. (2004), Atteia (2003) and
Firmani et al. (2006), may not be true but pseudo since they are
correlated with an unavoidable threshold effect (namely Malmquist
bias). In this case, some previous empirical relations will meet a
serious challenge on whether they are reliable for cosmological
applications or not.

Recently, Zhang et al. (2006a) put forward a new redshift/luminosity
estimator of  relative spectral lag (RSL, $\tau_{rel, 31}$), which
is defined as the ratio of spectral lags between light curves
observed in energy channels 1 and 3 to the full width at half
maximum ($FWHM$) of the light curve in channel 1. Based on analyzing
the RSL for 9 long BATSE GRBs with known redshift, they found that
the RSLs are also tightly correlated with the redshift or
luminosity, as follows:

\begin{equation}
logz=a-b\tau_{rel, 31}
\end{equation}
\begin{equation}
logL=\eta-\xi\tau_{rel, 31}
\end{equation}
where $a=1.56\pm0.24, b=9.66\pm1.86, \eta=55.44\pm0.63,
\xi=23.07\pm4.88$ and $\tau_{rel, 31}$ is normally distributed with
a mean value of $\mu=0.102$ and a standard error of $\sigma=0.045$.
The spearman rank-order correlation coefficients of the two
relations are -0.88 ($p\sim1.5\times10^{-3}$) and -0.83
($p\sim5\times10^{-3}$) respectively, indicating the RSL could be a
redshift/luminosity indicator (see Zhang et al. 2006a for the
details and Peng et al. 2007 for further studies). Among the nine
sources in Zhang et al. (2006a), eight redshifts are estimated from
the $L_p-E_{p}$ relation by Yonetoku et al. (2004) (assumed the
derived redshifts are comparable with the observed ones no matter
whether the $L_p-E_{p}$ relation is artificial). Strictly speaking,
the redshifts should be measured from spectroscopy rather than some
empirical relations. Meanwhile, we see the RSL is an unique and
intrinsic quantity since such definition can reduce both Doppler and
cosmological time dilation effects on the observations owing to
$\tau_{lag} \propto\Gamma^{-2}\propto(1+z)$ and $FWHM
\propto\Gamma^{-2}\propto(1+z)$ (Zhang et al. 2005, 2006b; Norris et
al. 2000; Kocevski \& Liang 2006). On the other hand, the RSL is
independent of energy bands because $\tau_{lag}$ and $FWHM$ are
roughly proportional to $E^{-0.4}$ respectively (e.g. Fenimore et
al. 1995; Norris et al. 1996; Reichart et al. 2001; Zhang et al.
2007).  In such case, the RSL is also independent of energy bands,
implying that the measurement of RSL is not influenced by different
instruments with distinct energy sensitivity.

Therefore, the primary task of this work is to check if the
luminosity/redshift-RSL (hereafter L/R-RSL) relation can pass the
so-called generalized test here. In addition, we investigate what we
can get from the RSL relations within its current precision level.
Finally, we focus our study on whether there are outliers violating
from the new L/R-RSL relation in statistics.

\section{Methodology}

We adopt the same method proposed by Schaefer \& Collazzi (2007) to
test the L-RSL relation. According to their unified expression, the
luminosity in our empirical relation can be written as

\begin{equation}
L={\cal A}[\kappa(1+z)^{Q}]^{m}
\end{equation}
where the parameters ${\cal A}$, $Q$, $m$ and $\kappa$ are usually
determined by empirical relations. In particular, the power index
$Q$ represents the cosmological time dilation effect to the
observation. We take $Q=0$ because the RSL already eliminated the
cosmological effect. By comparing Eq. (3) with Eq. (2), we let
$\kappa=10^{\tau_{rel, 31}}$ and rewrite the Eq. (3) as

\begin{equation}
logL=log{\cal A}+m\tau_{rel, 31}
\end{equation}
Combining Eqs. (2) and (4), we easily have ${\cal A}=10^{\eta}$ erg
s$^{-1}$ and $m=-\xi$.

On the other hand, the luminosity can also be obtained from the
following expression if the redshift and bolometric peak flux are
given

\begin{equation}
L=4\pi d_{l}^{2}{\psi}(1+z)^{-B}
\end{equation}
where $\psi$ and $B$ denote bolometric peak flux and cosmological
correction for time dilation, respectively. In this equation, we
have $\psi=P_{bolo}(z)$ and $B=0$ for a burst with measured spectral
parameters (namely $\alpha$, $\beta$ and $E_p$ for Band spectrum
(Band et al. 1993)). The luminosity distance $d_{l}$ is calculated
throughout this paper with respect to the unchanging parameters
$H_{0}$=100$h$ km s$^{-1}$Mpc$^{-1}$ ($h=0.71$), $\Omega_{k}=0$ for
a flat universe, $\Omega_{M}=0.27$ and $\Omega_{\lambda}=0.73$ (Dai
et al. 2004). Meanwhile, we suppose the unchanging cosmological
constant of $w=-1$ in advance and take the speed of light as
$c=3\times10^{5}$ km s$^{-1}$.

With given parameters value, after eliminating $L$ from Eqs. (3) and
(5), we derive the separate variable functions (SVFs) as follows

\begin{equation}
F(\kappa, \psi)=[(H_{0}/c)^{2}/4\pi]({\cal A}\kappa^{m}/\psi)
\end{equation}
\begin{equation}
F(z)=(H_{0}/c)^{2}d_{l}^{2}
\end{equation}
Here, note that Eq. (6) instead of Eq. (7) is determined by the
L-RSL relation. An ideal luminosity relation should be expected to
approach $F(\kappa,\psi)=F(z)$ in term of statistical principium.
Therefore, we use these relations to investigate whether a GRB
sample can be described by the L-RSL relation.

\section{Results}

In this section, we list the derivations from the L/R-RSL relation
and compare them with the observations. We list the results of the
generalized test for two different GRB samples. In particular, we
adopt another test to examine whether there are outliers violating
the L/R-RSL relation in comparison to other luminosity estimators.

\subsection{RSL Distribution}

Measuring the cosmological redshift of GRBs is important for
estimating the energy output, the distance and intrinsic parameters
including jet angles (e.g. Sari et al. 1999; Frail et al. 2001), the
Lorentz factor of ejection (e.g. Dermer et al. 1999; Panaitescu \&
Kumar 2002; Molinari et al. 2006; Zhang et al. 2006), the medium
density etc (see Piran 2005 and M\'{e}sz\'{a}ros 2006 for reviews).
A redshift distribution for GRBs has been adopted to calculate an
evolving star formation rate and to explore a GRB rate evolution
especially in the high redshift Universe (see Bromm \& Loeb 2002,
2006; Natarajan et al. 2005; Lloyd-Ronning, Fryer \& Ramirez-Ruiz
2002; Daigne et al. 2006; Le \& Dermer 2006; Salvaterra et al. 2007;
see also Coward 2007 for a review). However, it is difficult to
accurately determine the distribution because of present incomplete
samples (Tanvir \& Jakobsson 2007). Nonetheless, statistical
approach to the distribution is still important, for instance, to
probe the efficiency of GRB production (Daigne et al. 2006).

Here, using the measured redshifts we adopt Eq. (1) to estimate the
RSL of each burst in Schaefer's (2007) sample of 69 GRBs, including
35 pre-Swift and 34 Swift bursts, in order to check if the RSLs are
indeed normally distributed. This provides us a new way to test the
L/R-RSL relations shown in Eqs. (1) and (2).
\begin{figure}
\begin{center}
\centering \resizebox{8cm}{6cm}{\includegraphics{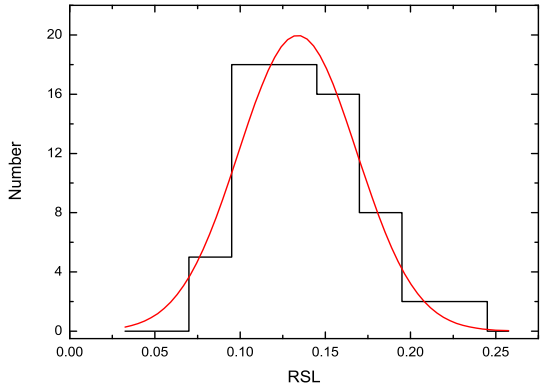}}
\caption{The distribution of RSLs for Schaefer's sample of 69 GRBs
(histogram). The solid curve represents a Gaussian fit to the data,
in which we achieve the standard deviation $\sigma=0.035$ and the
mean $\mu=0.13$ ($\chi^{2}/d.o.f.=1.4$).}
\end{center}
\end{figure}
Figure 1 shows the distribution of the RSLs together with the best
fit line of Gaussian curve. We see that the RSLs have a Gaussian
distribution with the standard deviation $\sigma=0.035$ and the mean
$\mu=0.13$, which are close to the corresponding values of
$\sigma=0.045$ and $\mu=0.102$ in Zhang et al. (2006a). The small
discrepancy comes from the statistical fluctuation and thus is not
significant. It is an interesting result since the different
instruments with distinguishing energy sensitivity can present the
same normal distribution of RSLs. The fact demonstrates that the RSL
is independent of energy bands, at least insensitive to energy
channels.

\subsection{Luminosity Function}

We already know the observed luminosity is usually calculated from
Eq. (5) through the measurements of redshift and peak flux for an
assumed cosmology. Apart from this, eliminating $\tau_{rel, 31}$
from Eqs. (1) and (2) one can estimate the luminosity using the
following relation
\begin{equation}
logL=\eta-\xi(a-logz)/b,
\end{equation}
where the luminosity only depends on the redshift measurement.
Besides, this relation allows us to conveniently estimate the
luminosity of each burst once its redshift has been measured. It
needs to be pointed out that both luminosity calculations are
relevant to a special cosmology. In theory, the empirical Eq. (8)
can match the observed values well if it is reliable to estimate the
luminosity.

Fortunately, large amount of high-quality data have resulted in the
determination of cosmological parameters with a rather good
precision (Balbi 2006). This means different luminosity relations
depend not too much on an assumed standard cosmology but on
themselves. We apply Eqs. (5) and (8) to the Schaefer's sample
respectively and compare the observed luminosities with the
estimated ones in Figure 2, from which we see that they both follow
an analogous trend with the increase of redshift, which supports the
result of Wei \& Gao (2003) obtained from the luminosity-variability
relation. In addition, we find both of them have a very close median
value, i.e., $1.89\times10^{52}$ erg s$^{-1}$ and
$1.64\times10^{52}$ erg s$^{-1}$ for the observed and estimated
luminosities, respectively. To evaluate the degree of consistency,
we put the lines of upper and lower limits on log$L$ (1 $\sigma$
confidence level) arising from the L-RSL relation. Note that
$\sigma\simeq[(\sum\limits_{i=1}^N(y_{o,i}-y_i))/(N-1)]^{1/2}$ whose
$y_{o,i}$ and $y_i$ respectively stand for the observational and
theoretical (or derivative) quantities.
\begin{figure}
\begin{center}
\centering \resizebox{8cm}{6cm}{\includegraphics{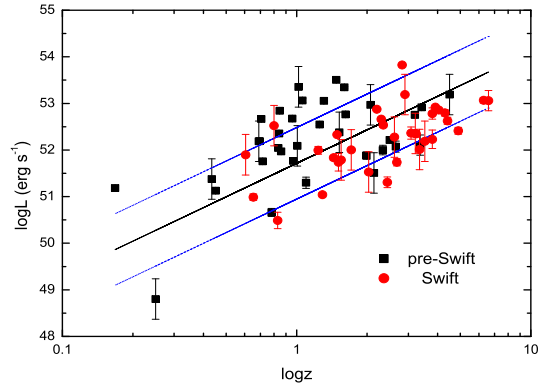}}
\caption{Estimated (solid line) versus observed (filled squares and
circles respectively denote for pre-Swift and Swift bursts)
luminosities. The upper and lower limits for L-RSL relation are
represented as dotted lines within 1$\sigma$ confidence level of
$\sigma\simeq0.76$.}
\end{center}
\end{figure}
As can be seen in figure 2, both pre-Swift and Swift bursts follow
the empirical luminosity relation robustly. Quantitatively, the
L/R-RSL relation can account for about 72\% sources in Schaefer's
sample within 1 $\sigma$ level. For larger confidence levels, say
3$\sigma$, no sources are found to violate the relation. Moreover,
we notice that most data points exist within the range of redshift
from $z\sim$0.5 to 4, indicating that the current GRB sample lacks
much more sources with lower and higher redshifts.

In Zhang et al. (2006a), if building a relation of $\tau_{rel, 31}$
with $1+z$, we can rewrite Eq. (1) as
$log(1+z)=(1.37\pm0.19)-(6.54\pm1.46)\tau_{rel, 31}$ with
probability $P=0.002$. Combining this relation with Eq. (2), one can
get a rough luminosity function of $L\propto(1+z)^{3.52\pm1.08}$.
The power law index is significantly different with that of either
$\sim1.4$ drawn by Lloyd-Ronning et al. (2002) from the
luminosity-variability relation or $\sim1.7$ gotten by Kocevski \&
Liang (2006) from the luminosity-lag relation. However, our
luminosity function is more close to $L\propto(1+z)^{2.70\pm0.60}$
up to $z\simeq6$, which is based on the suggestion that GRBs
perfectly matches the history of cosmological star-formation
(Hopkins 2004). This implies that the L/R-RSL relation might favor a
higher redshift estimation.

\subsection{SVF Curve and Violator Test}

From Eq. (7), we get the theoretical SVF curve for the measured
redshift (z) and the given cosmological parameters. Here, we
likewise let $z_{max}=20$ in order to make a direct comparison with
the Schaefer's results. Then one can acquire the maximum value $
F_{max}(z)\sim3002$  at $z=z_{max}=20$ (Bromm \& Loeb 2002).
\begin{figure}
\begin{center}
\centering \resizebox{8cm}{6cm}{\includegraphics{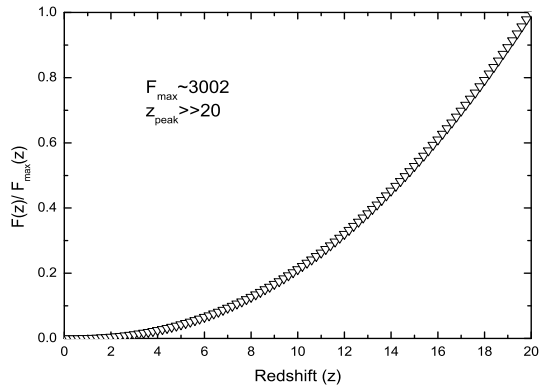}}
\caption{Normalized theoretical SVF ($F(z)/F_{max}$) curve for the
L-RSL relation.}
\end{center}
\end{figure}
Figure 3 shows the normalized SVF curve, from which we find it
behaves same as that of the $L-N_{p}$ relation plotted by Schaefer
\& Collazzi (2007). The reason for the agreement is that both RSL
and $N_{p}$ are intrinsic variables and uninfluenced by the
cosmological correction for time dilation. In evidence, the L-RSL
relation satisfies Li test because the SVF increases monotonically
with redshift in its reasonable range.

In general, we name some special sources violating one given law as
violators or outliers. The violators as a ruler in any effective
tests usually reflect the validity of anyone of the luminosity
relations. A source passes the Nakar \& Piran (2005) test means it
locates in the region of $F/F_{max}<1$. To examine if there are some
bursts violating the L-RSL relation, we apply this method to the two
different GRB samples: one is taken from Zhang et al. (2006a), in
which 9 single-pulsed bursts holding both long spectral lags and
wide widths are included (hereafter Zhang's sample); Another is
Schaefer's sample of 69 sources as mentioned above. The motivation
of this selection is to exclude the hopeless effect of sample
selection on our results. At the same time, we also need to diagnose
the degree of the derived data points according with the L/R-RSL
relation in a quite different manner.

We use the spectral indices ($\alpha$, $\beta$, and $E_{p}$)
presented in Zhang et al. (2006a) to calculate the bolometric peak
flux $P_{bolo}$ in the unit of erg cm$^{-2}$ s$^{-1}$ (where the
energy range $E_{min}=50\ {\rm keV}$ and $E_{max}=300\ {\rm keV}$ is
selected for the BATSE instrument; see Schaefer 2007 for the
details). With the measured $\tau_{rel, 31}$ and $P_{bolo}$ for 9
bursts, we obtain the SVF values of $F(\kappa, \psi)$ from Eq. (6)
and the corresponding deviations within 1$\sigma$ error, as shown in
Figure 4.
\begin{figure}
\begin{center}
\centering \resizebox{8cm}{6cm}{\includegraphics{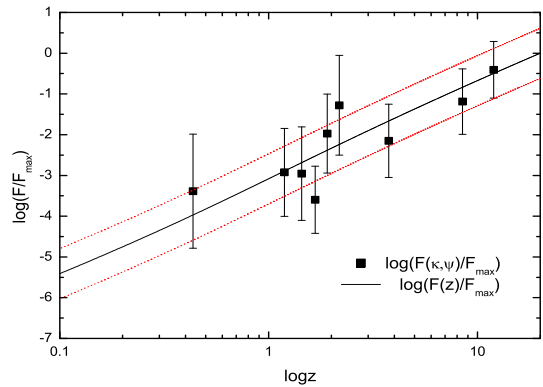}}
\caption{Comparison of the theoretical SVF ($F(z)/F_{max}$, solid
line) with the values for Zhang's sample ($F(\kappa, \psi)/F_{max}$,
filled squares) calculated from the L-RSL relation. The limits of 1
$\sigma$ ($\simeq0.62$) level are marked with the red dotted lines.}
\end{center}
\end{figure}
In addition, a theoretical SVF curve (Eq. (7)) that is proportional
to luminosity distance squared is plotted to compare with the data
points in the Zhang's sample. We find $\simeq33\%$ sources are
marginally present as violators at 1$\sigma$ level. It is also found
within a larger uncertainty limit (namely $\geq 2\sigma$) all the
data points are well consistent with the theoretical curve. It is
obviously seen that all the derived data points distribute in the
region of $log(F/F_{max})<0$, indicates no sources violate the Nakar
\& Piran test.

Now, we apply this test to the Schaefer's sample with the measured
redshift and $P_{bolo}$. Using Eq. (1), one can estimate the RSL for
each burst. As seen in figure 1, the derived RSLs are normally
distributed. Substituting the estimated RSLs to Eq. (6), one can
then obtain the $F(\kappa, \psi)$ values of 69 bursts.
Similarly, our L/R-RSL relation is also found to pass the Nakar \&
Piran test because all data points have evidently small values of
$F/F_{max}<1$ (Figure 5).
\begin{figure}
\begin{center}
\centering \resizebox{8cm}{6cm}{\includegraphics{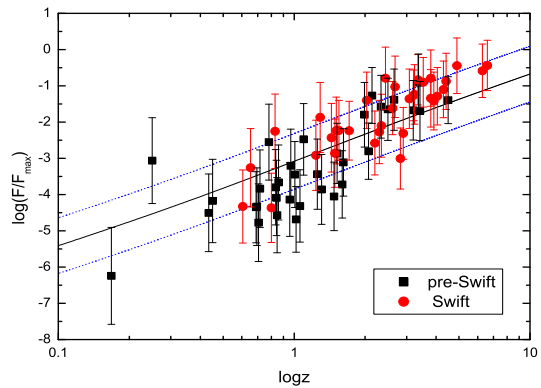}}
\caption{Comparison of the theoretical SVF ($F(z)/F_{max}$, solid
line) with the values for Schaefer's sample ($F(\kappa,
\psi)/F_{max}$, filled squares for the pre-Swift and filled circles
for the Swift bursts) estimated from the L-RSL relation. The limits
of 1 $\sigma$ ($\simeq0.77$) level are marked with the blue dotted
lines.}
\end{center}
\end{figure}
It is worthwhile to emphasize that within 1$\sigma$ errors 64\% data
points estimated from both pre-Swift and Swift bursts match the
theoretical curve well. That is to say, the fraction of potential
outliers for L/R-RSL relation is $\sim36\%$. However, a higher
confidence interval of 99.8\% (3$\sigma$) for the selective
Schaefer's sample reports that no violators deviate from the
theoretical line under the current precision level of the L/R-RSL
relation.

\subsection{Comparison with other estimators}

In principle, the real outliers should be basically the same for all
the unbiased luminosity estimators. Unfortunately, people usually
give different even opposite judgements on the same burst. For
example, some previous works claimed all but two bursts (GRB 980425
and GRB 031203) were consistent with the Amati relation (Amati et
al. 2002) as well as the Ghirlanda relation (namely the
$E_{p}-E_{\gamma}$ relation) (Ghirlanda et al. 2004; see also
Ghisellini et al. 2006). This is a very strong contrast with the
result of 88\% being outliers reported by Nakar \& Piran (2005) for
the Amati relation. According to Buter et al.'s (2007)
interpretations, the Amati relation is likely caused by threshold
effect and thus artificial. Panaitescu (2007) suggests the Ghirlanda
relation is probably a consequence of the Amati relation and can not
work for GRB 050416A. Rizzuto et al. (2007) find GRB 050416A is also
an outlier of luminosity-variability relation. In addition, Campana
et al. (2007) argues that at least five bursts (including GRB
050416A) are outliers of the $E_{p}-E_{\gamma}$ relation. Another
analysis shows it is not the case (Ghirlanda et al. 2007). The
outlier identification is a crucial target to judge the
practicability of any luminosity relations for cosmological studies.
However, the important topic is still controversial or uncertain for
most previous luminosity relations.

It is therefore necessary to examine if these potential outliers
(Table 1) exist in the same manner for the L/R-RSL relation. We
select 10 bursts in this study, in which 7 sources [GRB 050401 (8),
GRB 050416A (4), GRB 050603 (5), GRB 050922C (6), GRB 051109A (7),
GRB 060206 (10) and GRB 060526 (9)] are taken from the Schaefer's
sample of 69 sources except GRB 980425 (1), GRB 031203 (3) and GRB
060218 (2) (These bursts have been distinguished with the sequence
number in parenthesis). The last three long bursts hold comparably
lower luminosities and are associated with supernova. However, GRB
060218 unlike GRB 980425 and GRB 031203 is not an outlier with
respect to the Amati relation.

\begin{table*}
\centering \caption{Special bursts as the potential candidates of
``outlier'' for luminosity relations}
\begin{tabular}{lcccccll}
\hline\hline
GRB &Instrument & z&Log$L_{obs}$&Log$L_{der}$& Log[$F(\kappa, \psi)/F_{max}$] & Relation  & Ref.\\
\hline
(1)&(2)&(3)&(4)&(5)&(6)&(7)&(8)\\
\hline
 050401$^\ast$  &\textit{Swift}&2.9&53.19& 52.82&-2.31$\pm$0.72&$E_{p}-E_{\gamma}$&1\\
 050416A$^\ast$ &\textit{Swift}&0.653 &50.99&51.27&-3.26$\pm$1.08&$E_{p}-E_{\gamma}$&1, 2\\
 &&&&&& $L_{p}-V$&3\\
 050603         &\textit{Swift}&2.821 &53.83&52.79&-3.00$\pm$0.84&$E_{p}-E_{\gamma}$&1\\
 050922C$^\ast$ &\textit{Swift}&2.199 &52.87&52.53&-2.58$\pm$0.88&$E_{p}-E_{\gamma}$&1\\
 051109A        &\textit{Swift}&2.346&52.54&52.59 &-2.10$\pm$0.87&$E_{p}-E_{\gamma}$&1\\
 060206$^\ast$ &\textit{Swift}&4.05 &52.86&53.17&-1.29$\pm$0.79&$E_{p}-E_{\gamma}$&1\\
 060526$^\ast$ &\textit{Swift}&3.21 &52.35&52.92&-1.26$\pm$0.70&$E_{p}-E_{\gamma}$&1\\
 060218$^\star$ &\textit{Swift}&0.033 & 46.69& 48.18&-4.93$\pm$2.25&?&---\\
 980425$^\star$ &{\small \textit{BeppoSAX}}&0.0085&47.11&46.77&-7.42$\pm$2.62&$E_{p}-E_{iso}$&4, 5, 6\\
&&&&&& $E_{p}-E_{\gamma}$&4, 5, 6\\
 031203$^\star$ &{\small \textsl{INTEGRAL}}&0.106 &48.56&49.39&-4.53$\pm$1.93&$E_{p}-E_{iso}$&4, 5, 6\\
&&&&&& $E_{p}-E_{\gamma}$&4, 5, 6\\
\hline
\end{tabular}
\begin{flushleft}
\textit{Note}---Col. (1) and (3) are respectively the burst name and
redshift; Col. (2) stands for the exploring instrument; Col. (4) and
(5) represent the observed and derived logarithmic luminosities (erg
s$^{-1}$), respectively; Col. (8) denotes the references of the
outliers for luminosity relations marked in col. (7); Col (6) are
the derived SVF values in logarithmic form.
\\Refs.--- 1. Campana et al. 2007; 2. Panaitescu 2007; 3. Rizzuto et
al. 2007; 4. Ghisellini et al. 2006; 5. Ghirlanda, Ghisellini \&
Lazzati 2004; 6. Amati et al. 2007.
\\$^\ast$ Ghirlanda et al. (2007) argued these source were not
outliers for the $E_{p}-E_{\gamma}$ relation. \\$^\star$ The three
bursts are associated with supernova and belong to low luminosity
burst class. Unlike GRB 980425 and GRB 031203, GRB 060218 is however
in excellent agreement with the Amati relation (Ghisellini et al.
2006; Amati et al. 2007). Meanwhile, this burst has a similar energy
output with GRB 031203. Ghisellini et al. (2006) suggest GRB 980425
and GRB 031203 as a twin of GRB 060218 might be correlated with the
$L-E_p-T_{0.45}$ (Firmani et al. 2006). \end{flushleft}
\end{table*}
\begin{figure}
\begin{center}
\centering \resizebox{8cm}{6cm}{\includegraphics{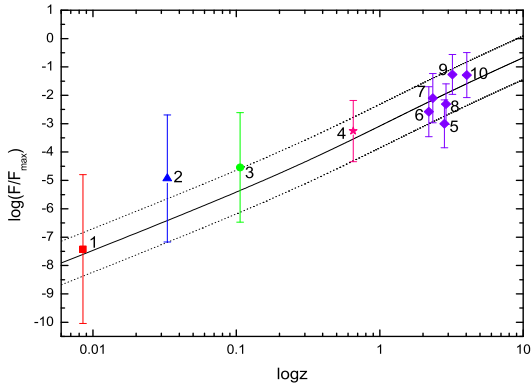}}
\caption{Examining the potential ``outliers'' violating some
previous luminosity relations in Table 1. The numbers represent the
10 potential outliers discussed by previous authors. Four bursts
with low luminosity are unusual and have been identified with
different color symbols (GRB 980425(1, square), GRB 061208 (2,
triangle), GRB 031203 (3, circle) and GRB 050416A (4, star)). The
solid and dotted lines are same as in Fig. 5.}
\end{center}
\end{figure}
Figure 6 shows within 1$\sigma$ level $\sim80\%$ bursts as outliers
of other luminosity relations perfectly follow the L/R-RSL relation.
We stress that GRB 050416A (marked with star) is a typical outlier
for both $L-V$ and $E_{p}-E_{\gamma}$ relations, but good consistent
with the L/R-RSL relation here. Surprisingly, we also see GRB 980425
fully matches and GRB 031203 is marginally consistent with the
relation within 1 $\sigma$ region. GRB 060218 is also found to
follow the L/R-RSL relation well within 3 $\sigma$ levels. These
potential outliers for other luminosity relations do not violate the
L/R-relation. The interesting results seem in turn to show the
L/R-RSL relation is potentially an expected cosmological tool.

\section{Discussion and Summary}

As mentioned in \S 3.3, the agreement between SVF curve in Figure 3
and that of the $L-N_{p}$ relation provides us an indirect clue that
the L-RSL relation is expectantly true. Likewise, we have shown in
Figure 4 that the 9 data points derived from the L/R-RSL relation
match well with the theoretical SVF curve, which in turn proves both
$L_p-E_{p}$ and L/R-RSL relations are equally reliable because the
data are actually related with the $L_p-E_{p}$ relation. Besides,
the R-RSL relation can be used to estimate redshift without
constraints by theoretical cosmological models.

Although many luminosity/redshift estimators have been constructed
so far to obtain either a synthetic redshift or luminosity distance,
it is still difficult to answer which one is the best. In theory, a
powerful estimator should be applicable to both lower and higher
redshift sources without producing an obvious evolutionary effect
for fitted parameters in a statistical point of view. Besides, the
empirical data points from this estimator should be distributed
along the smooth theoretical SVF curve shown in Figure 4. However,
some factors involving the systematic and measurement errors and the
gravitational lensing effects (Oguri \& Takahashi 2006) can cause
additional uncertainties that make the observations to deviate
largely from the expected curve. Another different effect is the
Malmquist biases (Schaefer 2007; Butler et al. 2007) that lead the
estimated data points to approach the SVF curve. Note that the
L/R-RSL relation is built with only 9 long bursts. To get more
precise RSL relations, calibration as well as an in-depth test
required, especially for many but several bursts with measured
redshifts.

We summarize our results as follows: (1) We have clarified that the
RSLs have a Gaussian distribution; (2) Our calculations for RSLs and
luminosities (or redshifts) are comparable with those of
observations, indicating that the L/R-RSL relation may be a
potential tool for cosmological study; (3) The behavior of
luminosity increasing with redshift confirms the result of Wei \&
Gao obtained from the luminosity-variability relation; (4) We have
tested the L/R-RSL relation for two different GRB samples and found
that there exist 36 per cent of Schaefer's sample are outliers
within 1$\sigma$ confidence level, but no violators at 3$\sigma$
regions within the current precision of L/R-RSL relation; and (5)
The potential outliers for other luminosity relations can match the
L/R-RSL relation well.

\section*{Acknowledgments}
We gratefully thank V. Bromm and D. Yonetoku for good
communications. We give thanks to Chul-Sung Choi and Heon-Young
Chang for helpful discussions. Z. B. Z would like to acknowledge
Korea Astronomy and Space Science Institute (KASI) for postdoctoral
fellowship.

\label{lastpage}

\end{document}